\documentclass[12pt,preprint]{aastex}

\newcommand{\myemail}{jlg@ujaen.es}

\slugcomment{{\sc Accepted for publication in ApJ Letters}}

\shorttitle{The Futile Search for Galactic Disk Dark Matter}
\shortauthors{Pesta\~{n}a and Eckhardt}

\begin{document}

\title{The Futile Search for Galactic Disk Dark Matter}

\author{Jos\'{e} Luis Garrido Pesta\~{n}a\altaffilmark{1}}
\affil{Departamento de F\'{\i}sica, Universidad de Ja\'en, Campus Las Lagunillas, 23071 Ja\'en, Espa\~{n}a}

\and

\author{Donald H. Eckhardt}
\affil{Canterbury, NH 03224-0021, USA}

\altaffiltext{1}{Electronic address: \myemail.}

\begin{abstract}
Several approaches have been used to search for dark matter in our galactic disk, but with mixed results: {\em maybe yes and maybe no}.  The prevailing approach, integrating the Poisson-Boltzmann equation for tracer stars, has led to more definitive results: {\em yes and no}.  The touchstone {\em yes} analysis of \citet{BFG:92} has subsequently been confirmed or refuted by various other investigators.  This has been our motivation for approaching the search from a different direction: applying the Virial Theorem to extant data.  We conclude that the vertical  density profile of the disk is not in a state of equilbrium and, therefore, that the Poisson-Boltzmann approach is inappropriate and it thereby leads to indefensible conclusions.
\end{abstract}


\section{Introduction}
\label{sect:in}

The mass distribution of our Galaxy can be derived  from a knowledge of forces both parallel and perpendicular to its midplane. Our knowledge of the parallel forces comes by inference from measurements of the generally flat rotation profiles of other spiral galaxies and by direct measurements of Milky Way stellar motions that are used to estimate galactic rotation. The conventional explanation for the observations is that  dark matter (DM) in a halo somehow ``conspires'' with visible matter to keep the rotation profiles flat.

Studies of forces perpendicular to the galactic midplane have also led to DM hypotheses, but this DM is confined to the disk where it purportedly has a higher density than the halo DM. The search for disk DM, 
started 95 years ago by \citet{O:15}, has a  controversial history.  It has lasted even longer than the 75 year futile search for Vulcan~\citep{BS97}.

\citet{K:22} was the first to apply the kinetic theory of gases to interpret star counts and estimate the gravitational field of the disk in the vicinity of the Sun. \citet{O:32} refined the theory and wrote the seminal paper on the topic. The gravitational potential can be estimated from observations of the midplane or column densities and the velocity distributions of all observable stars and gases, and it can independently be estimated from observations of limited sets of tracer stars.  A discrepancy between the two estimates may be indicative of  DM. Numerous papers were published following Oort's lead, culminating with the 
 disk model of \citet{BFG:92} (BFG). The data they used to determine the gravitational potential comprised 13 stellar components (excluding a very small spheroid component) plus four gas components. They found that their data supported the hypothesis that disk DM exists.  Yet the significance of this finding is debatable because it has not been confirmed on using  
 K giant data from the \emph{Hipparcos} catalogue \citep{KGBDvA:03,HF:04}. 
Specifically, \citet{HF:04} (HF) updated the BFG model and proffered  the most comprehensive attempt yet at modeling the galactic disk mass density in the solar vicinity. 
The HF model has  11 stellar components and the same four gas components as the BFG model, plus a thick disk constituent, but no disk DM.

Another approach  is the study of the thickness of the neutral
hydrogen component of the Milky Way.  \citet{K:03} and \citet{KDKH:07}  analyzed flaring data in a Galaxy model with 6 gaseous and 4 stellar components (including a central bulge and bar).  To explain the observations, they required DM in a thick disk and in a ring - plus the usual DM halo.
Using a computer simulation of the evolution of galaxies, \citet{RLAD:08} concluded that our Galaxy includes disk DM.  Conversely, analyses of gravitational microlens observations toward the galactic center \citep{Petal:05,Hetal:06} found no evidence for DM in the disk; and, alternatively, the semblance of a similar, yet fictitious, DM disk is predicted by MOND \citep{M:01}.
Disk DM, like halo DM, is a broadly accepted working hypothesis, but it remains just a hypothesis that demands further investigation.  Independently being able to confirm or deny the existence of a DM disk would be important to fundamental physics and cosmology, so we have devised  a technique to analyze  
 the $z$ and $v_z$ distributions of Milky Way stars and gases by applying the Virial Theorem (Section \ref{vth}) to the HF disk model (Section \ref{hfdata}).

\section{Theory}
\label{th}

\subsection{The Virial Theorem}
\label{vth0}

We adopt the BFG toy model for which
 the mass density  of a uniformly flat Milky Way disk is  $\rho(z)=\rho(-z)$.  In the vicinity  of the Sun, at distance $r=r_{\odot}$ from the Galaxy rotation axis,
 the disk mass surface density (also called the column density) is
$$\Sigma=\int_{-\infty}^{\infty}\rho(z)\,dz;$$
 its kinetic energy surface density is
\begin{equation}
T=\Sigma\, \sigma^2/2
\label{eq:T}%
\end{equation}
(which, in effect, defines the variance $\sigma^2$ of the vertical velocity $v_z$);
 and its
  gravitational energy surface density is
\begin{equation}
V=-\int_{-\infty}^{\infty}\rho  \Psi'\,z\,dz=\int_{-\infty}^{\infty}\rho gz\,dz,
\label{eq:V1}%
\end{equation}  
where $\Psi$ is the gravitational potential and $g$ is the gravitational specific force (acceleration).  
For $(z/r_{\odot})^2\ll 1 $, the gravitational energy between horizontal layers parallel to the midplane is effectively independent of the gravitational energy of the orbit about the Galaxy rotation axis.  Then if the disk does not secularly disperse, the  Virial Theorem~\citep{HG:59,C:78} applies: 
 \begin{equation}
\overline{2T+V}=0,
\label{eq:vir}%
\end{equation}
where the overbar is the time averaging operator.  If the vertical mass column is in a state of equilbrium [ $\dot{\rho}(z)=0$ ],  then the instantaneous value of $2T+V$ is $0$.  If instead the mass column is not in a state of equilibrium, then there is an interchange of energy density back and forth between $T$ and $V$, and consequently $2T+V$ oscillates about $0$. 

\subsection{The Poisson-Boltzmann Equation}
\label{pbe}

For a disk vertical mass column that is in an  equilbrium state, the Poisson equation applies:
\begin{equation}
\Psi''(z)=4\pi G\rho(z),
\label{eq:Poisson}%
\end{equation}
where $G$ is the gravitational constant.  The boundary conditions are $\Psi(0)=\Psi'(0)=0$.
If an assembly of disk stars is isothermal, then $\sigma^2$ is independent of $z$, and if it is not isothermal, then it is composed of
multiple sub-assemblies
 that are isothermal.  The density distribution of all disk stars and gas clouds,
 ultimately partitioned among $N$ isothermal components,
 is the sum of $N$ Boltzmann distributions,
 \begin{equation}
        \rho(z)=\sum_{n=1}^{N}\rho_{n}(z)=\sum_{n=1}^{N}\rho_n(0)
        \, \exp \, [-\Psi(z)/\sigma_{n}^{2}].
        \label{eq:Boltzmann}%
  \end{equation}
Combining the Poisson (\ref{eq:Poisson}) and Boltzmann (\ref{eq:Boltzmann}) equations gives the
Poisson-Boltzmann equation,
\begin{equation}
\Psi''=4\pi G 
\sum_{n=1}^{N}
\rho_{n}\,\exp \, [-\Psi/\sigma_{n}^{2}].
\label{eq:full}%
\end{equation}

Spitzer derived an analytic solution
for the Poisson-Boltzmann equation with a single ($n=1$) isothermal
component,
\begin{equation}
        \Psi''=4\pi G\rho(0)
        \exp \,[-\Psi/\sigma^{2}]=2k^2\sigma^2\exp \,[-\Psi/\sigma^{2}]
        \label{eq:simple1},%
\end{equation}
where 
\begin{equation}
k^2\sigma^2=2\pi G\rho(0).
\label{eks}%
\end{equation}
Multiply both sides of Eq.~\ref{eq:simple1} by $\Psi'$, and then integrate, 
\begin{equation}
        (\Psi')^{2}=4k^2\sigma^4
                [1-\exp \, (-\Psi/\sigma^{2})].
           \label{eq:simple2}%
\end{equation}
Define $$u^{-2}=\rho(z)/\rho(0)=\exp(-\Psi /\sigma^{2}\,);$$
then $$\Psi=2\sigma^2\,\ln u,$$
so the LHS of Eq.~\ref{eq:simple2} is
\begin{equation}
(\Psi')^{2}=4\sigma^4 (u'/u)^2,
\label{lhs}%
\end{equation}
and the RHS of Eq.~\ref{eq:simple2} is 
\begin{equation}
4k^2\sigma^4 [1-\exp \, (-\Psi/\sigma^{2})]=4k^2\sigma^4[1-u^{-2}].
\label{rhs}%
\end{equation}
Equating Eqs.~\ref{lhs} and \ref{rhs},
\[
\frac{du}{\sqrt{u^{2}-1}}=
k dz.
\]
Integrating this equation from $u(0)=1$ to $u(z)$ gives $u(z)=\cosh kz$, and then the Spitzer
solution,
\begin{equation}
\rho(z)=\rho(0) u^{-2}=\rho(0)\;\mbox{sech}^{2} kz \, ,
\label{sprho}%
\end{equation}
and
\begin{equation}
\Psi(z)=2\sigma^{2}\ln u=2\sigma^{2}\ln \cosh kz \, .
        \label{sppsi}%
\end{equation}
This solution conforms with the steady state Virial Theorem ($2T+V=0$) because,  
by Eqs.~\ref{sprho} and \ref{eq:T},
\begin{equation}
\Sigma=\rho(0)\int_{-\infty}^{\infty}\,\mbox{sech}^{2} kz\,dz=2\rho(0)/k,
\label{sigsp}%
\end{equation}
\begin{equation}
2T=2\rho(0)\sigma^2/k;
\label{lhv}%
\end{equation}
and, by Eqs.~\ref{sprho}, \ref{sppsi}, \ref{eq:V1}, and \ref{lhv},
\begin{eqnarray}
V&=&-2\rho(0)\sigma^2\int_{-\infty}^{\infty}\mbox{sech}^{2} kz\;\frac{d(\ln \cosh kz)}{dz}\,z\,dz \nonumber \\
&=&-2\rho(0)\sigma^2\int_{-\infty}^{\infty}\mbox{sech}^{2} kz\,\tanh kz\,kz\,dz=-2\rho(0)\sigma^2/k
=-2T.
\label{rhv}%
\end{eqnarray}

We are aware of no analytic solution to Eq.~\ref{eq:full} for $N>1$, but it can be integrated numerically to find a unique, steady state  solution.  [This is not a trivial point because no such solutions can exist for 3-D Poisson-Boltzmann integrations such as modeling the structure of an isothermal star~\citep{PE:07}.]   We are accordingly poised to examine the problem by using the Virial Theorem.  The Virial Theorem approach has an advantage over Poisson-Boltzmann integration
because it 
merely requires that $\overline{2T+V}=0$, so an oscillatory solution is feasible, whereas the Poisson-Boltzmann integration requires that $2T+V=0$ and so only a time invariant solution is allowed.

\subsection{The Vertical Structure of the Disk}
\label{vth}

An infinitesimal layer of matter, $\rho(z)\,dz$, is the source of a specific force 
$$g=2\pi G\rho(z)\,dz$$
 in the direction of the layer.  Thus the gravitational energy surface
  density between infinitesimal layers of matter at
$z$ and $z+a$ is
$$-ga\,\rho(z+a)=-2\pi Ga\,\rho(z)\,\rho(z+a)\,dz,$$
and the gravitational energy surface
 density due to all layers that are separated by distance $a$ is
$$-2\pi Ga\int_{-\infty}^{\infty}\rho(z)\rho(z+a)\,dz=-2\pi G\Sigma^2a\int_{-\infty}^{\infty}p(z)p(z+a)\,dz       =-2\pi G\Sigma^2 q(a)a,$$
where $p(z)$ is the frequency function
\begin{equation}
p(z)=\rho(z)/\Sigma,
\label{eqpz}%
\end{equation}
and $q(a)$ is the frequency function  given by the convolution
\begin{equation}
q(a)= \int_{-\infty}^{\infty}p(z)\,p(z+a)\,dz.
\label{eq:conf}%
\end{equation}
The total gravitational energy surface density is then
\begin{equation}
V=-2\pi G\Sigma^2\int_{0}^{\infty}q(a)a\,da.
\label{eq:V}%
\end{equation}
The lower limit of this integral is $0$ because $a\ge 0$ to avoid double counting.

Combining Eqs.~\ref{eq:T}, \ref{eq:vir} (not time averaged), \ref{eqpz} and \ref{eq:V} gives
\begin{equation}
\frac{\rho(0)\sigma^2}{\pi G\Sigma^2}=2p(0)\int_0^{\infty}q(a)a\,da.
\label{fineq0}%
\end{equation}
Since both sides of Eq.~\ref{fineq0} are dimensionless, they need not have the same units of length. Then for computational convenience we implicitly define the unit of length on the RHS (and equations leading to it) by setting
\begin{equation}
p(0)=1/2,
\label{kdef}%
\end{equation}
so Eq.~\ref{fineq0} becomes the Virial theorem criterion for a state of equilibrium,
\begin{equation}
R=Q,
\label{qr}%
\end{equation}
where
\begin{equation}
R=\frac{\rho(0)\sigma^2}{\pi G\Sigma^2}\,,
\label{fineq1}%
\end{equation}
and
\begin{equation}
Q=\int_0^{\infty}q(a)a\,da.
\label{fineq}%
\end{equation}

In Eq.~\ref{fineq1}, $\Sigma/\rho(0)$ and $\sigma^2/G\Sigma$ retain the dimension pc.  
Using Eqs.~\ref{eks} and \ref{sigsp}, this ratio for the Spitzer solution is
\begin{equation}
R_{S}=1/2.
\label{rs}%
\end{equation}

We have examined various trial frequency functions, $p(z)$, using them in Eqs.~\ref{eq:conf} and \ref{fineq}  to evaluate  $Q$. These functions are not arbitrary.  Each must be bell-shaped: an even function of $z$ (that is, $p(z)=p(-z)$), with a single peak at $p(0)=1/2$ (that is, $zp'(z)\le 0$) and, of course,
$\int_{-\infty}^{\infty}p(z)\,dz=1.$
Moreover, we assume that 
$p(z)$ and all its derivatives are continuous, and that
 $p(z)$ is not a pathological frequency function, meaning that its $n$-th moment $\int_{-\infty}^{\infty}z^np(z)\,dz$ exists. 
Therefore, for large $z$, $p(z)\sim\exp(-\alpha z^{2m})$, where $\alpha$ is a positive constant and $m\ge 1$ is an integer.  With these provisos, it turns out that for all plausible frequency functions,
 \begin{equation}
 Q\approx 1/2.
\label{qap}%
\end{equation}
We offer two simple examples (for which $m=1$):

\begin{itemize}
\item For the Spitzer function (see Eqs.~\ref{sprho}, \ref{eqpz} and \ref{kdef}, with $k=1$), 
\begin{equation}
p_S(z)=\frac{1}{2}\,\mathrm{sech}^2 z,
\label{eq:spf}%
\end{equation}
which applies exactly for a single component model, but still might be a suitable approximation for a multiple component model.  For the Spitzer function,
\begin{equation}
Q_S= 1/2,
\label{eq:qsp}%
\end{equation}
so, as expected (see Eq.~\ref{rs}),
$R_S=Q_S$.

\item For the Gaussian frequency function,
\begin{equation}
p_G(z)=\frac{1}{2}\exp\left[-\pi\left(\frac{z}{2}\right)^2 \right],
\label{eq:spg}%
\end{equation}
for which
\begin{equation}
Q_G=\sqrt{2}/\pi=0.45.
\label{eq:qga}%
\end{equation}
\end{itemize}
These are merely examples.  The exact value for $Q$ is not important for this study.  Eq.~\ref{qap} is entirely adequate to restate the condition for equilibrium, Eq.~\ref{qr}, as
\begin{equation}
R=Q\approx 1/2.
\label{rx}%
\end{equation}

\section{Disk Mass Models}

\subsection{The HF Model}
\label{hfdata}

The 15 components of the HF disk mass model were itemized by \citealt{Fetal:06}.   From these, we calculate
\begin{equation}
\Sigma=\sum_1^{15}\Sigma_i=49.3\,{\mathrm pc^{-2}}\, M_{\odot},
\label{eq:surfd}%
\end{equation}
\begin{equation}
\pi G\Sigma^2=\pi\times 49.3^2\,{\mathrm pc^{-4}}\; (GM_{\odot})\,M_{\odot}
=34.4\times 10^{-27}\,{\mathrm pc^{-1}s^{-2}}\, M_{\odot},
\label{eq:lhs}%
\end{equation}
\begin{equation}
\sigma^2=\sum_1^{14}(\Sigma_i/\Sigma)\sigma_i^2=587\, \mathrm{km^2\,s^{-2}
=614\times 10^{-27}\,pc^2\,s^{-2}},
\label{eq:rhs}%
\end{equation}
\begin{equation}
\rho(0)=\sum_1^{15}\rho_i(0)=0.0914\, {\mathrm pc^{-3}}\,M_{\odot},
\label{eq:rho}%
\end{equation}
\begin{equation}
\rho(0)\sigma^2=56.1\times 10^{-27}\,{\mathrm pc^{-1}s^{-2}}\,M_{\odot},
\label{eq:rhs2}%
\end{equation}
and (see Eq.~\ref{fineq1}),
\begin{equation}
R_{HF}= 1.63,
\label{numeq}%
\end{equation}
which is too large compared with $Q\approx 1/2$ for an equilibrium solution to exist.
Hence, a Poisson-Boltzmann integration using the HF model, or any other such model, is an inappropriate and futile exercise.  Nevertheless, HF performed this integration  without disk DM and found the result to be compatible with vertical density profiles of K giant tracers.  We look at the same data and conclude that the HF model can be compatible with a Poisson-Boltzmann steady state solution only if disk DM exists.
For example, suppose that DM has the same vertical profile and $v_z$ variance as visible matter, but that visible matter accounts for only 30\%  of the total surface mass density.  Then $\rho(0)$ and $\Sigma$ increase by the factor $10/3$,  and thus $R_{HF}$ is modified to $1.63\times 3/10=0.49$, which is a plausible value for $Q$.  However, the revised model no longer satisfies the Poisson-Boltzmann equation and the change increases the vertical accelerations of the tracer stars by the factor $10/3$ and decreases their scale height by 70\%, and such a model conflicts drastically with K giant vertical profiles.  Introducing DM solves one problem but creates another.
 The overall problem can only be resolved by abandoning the requirement that the disk $z$ profile is in a steady state or that gravity is Newtonian.

\subsection{The BFG Models}
\label{bfgdata}

BFG conceived and examined ten disk models, nine of them containing varying formulations with DM.
Their Table 4 lists their observed model in Row 1 [$\rho_{1}(0)=0.1026\, {\mathrm pc^{-3}}\,M_{\odot}$ and $\Sigma_{1}=49.8\,{\mathrm pc^{-2}}\, M_{\odot}$], and their  ``best fit'' model (observed plus dark matter) in Row 3 [$\rho_{3}(0)=0.2596\, {\mathrm pc^{-3}}\,M_{\odot}$ and $\Sigma_{3}=83.9\,{\mathrm pc^{-2}}\, M_{\odot}$].    For the Row 1 model, the ratio of $\rho_{1}(0)/\Sigma_{1}^2$ to $\rho_{HF}(0)/\Sigma_{HF}^2$  is $(0.1026\times 49.3^2)/(0.0914\times 49.8^2)=1.100$, and for the Row 3 model, the ratio of $\rho_{3}(0)/\Sigma_{3}^2$ to $\rho_{HF}(0)/\Sigma_{HF}^2$  is $(0.2596\times 49.3^2)/(0.0914\times 83.9^2)=0.981$. If $\sigma^2$ does not change between the HF and BFG models, $R_1= 1.79$ and $R_3=1.60$.  Like $R_{HF}= 1.63$, $R_1$ and $R_3$ are incompatible with $R=Q\approx 1/2$. 

\section{An Oscillating Disk}
\label{od}

A Poisson-Boltzmann solution is feasible only if $2T+V=0$, but this requirement is incompatible with observations; compelling a steady state solution is tantamount to compelling the existence of DM or non-Newtonian gravity.
Suppose instead that a disk column's vertical profile varies periodically with time.  Then $2T+V$ oscillates about $\overline{2T+V}=0$,  so that at any epoch, $2T+V$ could be substantially greater or less than zero.  In other words, $R$ oscillates about $Q$.  We use this broadened aspect to interpret the HF data, and uncover interesting possible ramifications.  
  
  The analysis so far has been quantitative, but from here on it will be descriptive and speculative - and accordingly less defensible.
Using dimensional analysis, we reckon that the $z$ oscillation period is of the order 
$[G\rho(0)]^{-1/2} = 50$ million years.
The estimated period
 is roughly a quarter of a galactic year, but this estimate is inexact because it is based on dimensional considerations only.  It would have been larger if we had used the mean midplane density $\overline{\rho(0)}$ over one cycle instead of the current midplane density (which we believe to be near the maximum value), and there undoubtedly are dimensionless factors that would arise  in a precise analysis, decreasing the estimated period by as much as a factor of 10, but increasing it no more than fourfold 
(each disk star orbiting the galactic center has a retrograde precession caused by the other disk stars, so the $z$ oscillation period must be less than the orbital period).  What we find puzzling about the stellar orbits is the cause of the symmetric clustering of their ascending and descending nodes. 

 Disk column oscillations are exchanges between gravitational, $V\propto -\Sigma^3/\rho(0)$, and kinetic, $T\propto \Sigma\sigma^2$,   energy surface densities.  Since $\Sigma$ does not vary with time, oscillations in $\rho(0)$ entail oscillations in $\sigma^2$; and the $\rho(0)$ and $\sigma^2$ variations are in phase, so that the peaks in midplane mass and kinetic energy densities coincide. 
 Eq.~\ref{numeq} tells us that at the current oscillation phase $\rho(0)\sigma^2$ is about three times higher than its mean value, but it does not tell us how close $\rho(0)\sigma^2$ is to its peak value.  At the peak, there would be no net flow of stars toward or away from the midplane.  Unaware of any evidence for such a flow, we suppose that  $\rho(0)\sigma^2$ is close to its peak, and corroborate this with data from Table A.1 of \citet{Fetal:05}, selecting single giants (* tag in Column 24)  in the {\em smooth background} ($g=3$ in Column 34) that
  are in the two zones defined by $100 \le |z|\le 150$ pc.
There are 467 stars in the northern zone and 224 stars in the southern zone.  We calculate that the two zones  are approaching each other at  $1.3\pm 1.4$ km s$^{-1}$.  The Virial Theorem requires an oscillation with substantial energy flow, but these numbers tell us that there is negligible net mass flow now (at least for the giants), so the present epoch appears to be one of maximum $\rho(0)\sigma^2$.
With little net mass flow toward or away from the midplane, the $v_z$ velocity distributions may appear to be isothermal, but   at epochs of maximum flow, the net flow velocities would be of the order
 $\Sigma[G/\rho(0)]^{1/2}=10\,\mathrm{km\,s^{-1}}$.  Considering its imprecision, this is not much different from $\sigma_{HF}=24\,\mathrm{km\,s^{-1}}$.  A net outward flow would be obvious in about 10 million years. 

We extend our $z$ coordinate system  to a standard  cylindrical coordinate $(r,\vartheta,z)$ system. Then  we expect 
 oscillations elsewhere on the disk with periods proportional to $\overline{\rho(r,\vartheta,0)}^{\;-1/2}$. Because these mean midplane densities vary, so do the $z$ oscillation periods.
Just small differences in $\overline{\rho(r,\vartheta,0)}$ between neighboring disk columns can effect large changes in their relative $z$ oscillation phases and, therefore, large differences in their instantaneous midplane densities  $\rho(r,\vartheta,0)$. 

The increased midplane mass and kinetic energy densities caused by oscillations are conducive to the formation of stars, including the massive O and B stars whose luminosity lifetimes would be relatively ephemeral compared with the disk oscillation periods.  We therefore conjecture that the disk spirals are where the midplane mass and kinetic energy densities are now high, so stars are being formed there at 
relatively high rates.
  Vertical oscillations also occur in the gaps between the spirals, but their midplane mass and kinetic energy densities are now low (and their column gravitational energies are high) because of phase differences, so brighter stars that were created when their midplane mass and kinetic energy densities  had been high have sinced waned.  A computer simulation might tell us the shape that $p(z)$ takes in the gaps, for we do not even know whether the frequency function remains unimodal.
But neither do we know whether our conjecture concerning the formation of spirals is correct; nor, indeed, can we be sure that the disk actually oscillates, for conceivably the HF data  can be explained with a non-Newtonian gravitational theory.  However, we maintain that the conventional Poisson-Boltzmann interpretation of the data is demonstrably wrong, and we offer the disk oscillation hypothesis as one plausible  alternative.  

\section{Conclusion and Implications}
\label{ci}

We have used the Virial Theorem to study the problem of determining the density of matter near the Sun because it is less restrictive than the Poisson-Boltzmann approach. In fact, we have shown that all the comparisons of tracer star densities with Poisson-Boltzmann models or with equilibrium models in general, for example \citet{K:22,O:32,BFG:92,K:03,HF:00,HF:04}, have been wrong. The reason is that the 
disk is not in a state of equilibrium ($2T+V\ne 0$).  Furthermore, any model made compatible with $2T+V=0$ by adding DM will then be incompatible with Poisson-Boltzmann integration models, and with observed tracer star densities.  No evidence can be found for DM by using this approach. 
Explaining the evidence (e.g., the BFG, \citet{K:03} and HF disk models, plus tracer star densities) requires disk oscillations or, perhaps, non-Newtonian gravity.  A cursory inquiry into the possibility of disk oscillations without DM leads to intriguing hints concerning the structure and dynamics of a galactic disk.

We contend that any astrophysical argument predicated on the existence of a DM disk - for example \citet{S:98,S:99,dBSZGK:05,KDKH:07,RPCB:09,GG:01} - does not warrant consideration. We reject the recently proposed model for the formation of disk galaxies within the $\Lambda$CDM framework because it would produce a DM disk \citep{RLAD:08,PBK:09}.  Finally, there is no fictitious semblance of a DM disk as predicted by MOND \citep{M:01,BFZA:09},  so our analysis argues against MOND.

\acknowledgments
We thank the referee for positive comments and constructive suggestions, and especially for drawing out attention to a key numerical error and encouraging us to resolve its effect.

\end{document}